\documentclass[useAMS,usenatbib,aas_macros]{mn2e} 
\usepackage {graphicx,subfigure,times,aas_macros}

\newcommand{\se}[1]{\S\ref{sec:#1}}

\newcommand{\Fig}[1]{Figure~\ref{fig:#1}}

\newcommand{\tab}[1]{Table~\ref{tab:#1}}
\newcommand{\be}{\begin{equation}}
\newcommand{\ee}{\end{equation}}
\newcommand{\bea}{\begin{eqnarray}}
\newcommand{\eea}{\end{eqnarray}}

\newcommand{\msun}{{\rm M}_\odot}
\newcommand{\Msun}{M_\odot}

\newcommand{\ifm}[1]{\relax\ifmmode#1\else$\mathsurround=0pt #1$\fi}
\newcommand{\kms}{\ifmmode\,{\rm km}\,{\rm s}^{-1}\else km$\,$s$^{-1}$\fi}

\newcommand{\kpc}{\,{\rm kpc}}

\newcommand{\Gyr}{\,{\rm Gyr}}

\newcommand{\ltsima}{$\; \buildrel < \over \sim \;$}
\newcommand{\lsim}{\lower.5ex\hbox{\ltsima}}
\newcommand{\gtsima}{$\; \buildrel > \over \sim \;$}
\newcommand{\gsim}{\lower.5ex\hbox{\gtsima}}

\def\la{\langle}

\def\sy{\,M_\odot\, {\rm yr}^{-1}}

\def\cms{\,{\rm cm}^{-2}}
\def\ergs{\,{\rm erg}\,{\rm s}^{-1}}

\def\Ms{M_*}

\newcommand{\Halpha}{H${\alpha}$}

\usepackage{color}

\begin{document} 

\large 

\title[metallicity drops]
{Gas Inflow and Metallicity Drops in Star-forming Galaxies}

\author[Ceverino et al.]
{\parbox[t]{\textwidth} 
{ 
Daniel Ceverino$^{1,2}$\thanks{E-mail: daniel.ceverino@cab.inta-csic.es},
Jorge S{\'a}nchez Almeida$^{3,4}$, 
Casiana Mu{\~n}oz Tu{\~n}{\' o}n$^{3,4}$, \\
Avishai Dekel$^5$, Bruce G. Elmegreen$^6$,  Debra M. Elmegreen$^7$, Joel Primack$^8$}
\\ \\
$^1$Centro de Astrobiolog{\'i}a (CSIC-INTA), Ctra de Torrej{\'o}n a Ajalvir, km 4, E-28850 Torrej{\'o}n de Ardoz, Madrid, Spain \\
$^2$Astro-UAM, Universidad Autonoma de Madrid, Unidad Asociada CSIC, E-28049 Madrid, Spain \\
$^3$Instituto de Astrof{\'i}sica de Canarias, E-38205 La Laguna, Tenerife, Spain \\
$^4$Departamento de Astrof\'\i sica, Universidad de La Laguna, E-38205 La Laguna, Tenerife, Spain \\
$^5$Center for Astrophysics and Planetary Science, Racah Institute of Physics, The Hebrew University, Jerusalem 91904, Israel \\
$^6$IBM Research Division, T.J. Watson Research Center, 1101 Kitchawan Road, Yorktown Heights, NY 10598, USA \\
$^7$Department of Physics and Astronomy, Vassar College, Poughkeepsie, NY 12604, USA \\
$^8$Department of Physics, University of California, Santa Cruz, CA, 95064, USA 
}

\date{}

\pagerange{\pageref{firstpage}--\pageref{lastpage}} \pubyear{0000}

\maketitle

\label{firstpage}

\begin{abstract}

Gas inflow feeds galaxies with low metallicity gas from the cosmic web, sustaining star formation across the Hubble time. 
We make a connection between these  inflows and metallicity inhomogeneities in star-forming galaxies, by using synthetic narrow-band images of the \Halpha \ emission line from zoom-in AMR cosmological simulations of galaxies with stellar masses of $\Ms \simeq 10^9 \ \Msun$ at redshifts $z=2-7$.
In $\sim$50\% of the cases at redshifts lower than 4, the gas inflow gives rise to star-forming, \Halpha-bright, off-centre clumps.
Most of these clumps have gas metallicities, weighted by \Halpha \ luminosity, lower than the metallicity in the surrounding interstellar medium by $\sim$0.3 dex, consistent with observations of chemical inhomogeneities at high and low redshifts.
Due to metal mixing by shear and turbulence, these metallicity drops are dissolved in a few disc dynamical times.
Therefore, they can be considered as evidence for rapid gas accretion coming from cosmological inflow of pristine gas. 

\end{abstract}


\section{Introduction}
\label{sec:intro}

A number of observational works have detected chemical inhomogeneities in the gas of
star-forming galaxies, both at high redshift \citep{Cresci10,Queyrel12} and in the local 
universe
 (Richards et al. 2014; see S{\'a}nchez Almeida et al.
2014a for a review).
Most galaxies develop negative metallicity gradients, with the metallicity decreasing 
inside-out  \citep{Edmunds95, vanZee98, Magrini07, Califa14,Stott14,Kehrig15}. 
On top of such large-scale variation, many galaxies show 
localized, kpc-size starbursts with the metallicity in the burst lower than in the surrounding 
interstellar medium 
\citep{SA13, SA14a,SA15}.
These metallicity drops are of the order of 0.3 dex or larger.
Since the time-scale for gas mixing in disk galaxies is expected 
to be short, on the order of the rotational period 
or shorter \citep{deAvillezMacLow02,YangKrumholz12,PetitKrumholz15}, 
the low-metallicity gas sustaining the starbursts must have arrived to the galaxies recently.  

A plausible interpretation of these observations is the 
triggering of star-formation by gas recently accreted from the cosmic 
web \citep{Cresci10, SA13, SA15}. Numerical simulations predict that the inflow of 
cosmic-web gas has sustained star formation in galaxies along the Hubble time  
\citep[][and references therein]{Dekel13}.  
The gas being accreted is metal-poor \citep{vandeVoortSchaye12}  and it is expected to 
induce giant star-forming clumps in early disks \citep{DSC, CDB, Mandelker14}.
The gas from the web may 
trigger a starburst as it gets compressed when approaching the disk,  or it builds up the 
gas mass in the disk which eventually gives rise to starbursts due to internal instabilities. 
Only if the buildup and triggering  occurs rather fast, in a time-scale similar to or shorter than the mixing 
time-scale, the recently accreted, low-metallicity gas can still dominate the mass of the star-forming clumps. 

This paper studies whether this conjecture is 
valid in cosmological simulations of galaxies, namely whether metal-poor star-forming gas clumps in high-z protogalaxies  can be associated with
 individual gas accretion events.  \citet{Verbeke14} simulate the effect of gas clouds falling 
into disks of various masses and types. Their conclusion is that gas accretion may
give rise to low metallicity starbursts depending on details of the accretion 
event, such as the accreted gas mass, the impact parameter of the stream, or whether 
the orbit is prograde or retrograde. These simulations are idealized with 
ad-hoc initial conditions. We want to see whether the phenomena of metallicity drops 
associated with cosmic gas accretion appear in a cosmological setup,
where gas infall is not imposed but appears naturally, as a result of the formation of large-scale structures.
 
We aim at studying and
characterizing the physical process, rather than at explaining a particular set of 
observations.   
However, we want to make predictions that connect a given observable such as the gas metallicity weighted by \Halpha \ luminosity, with cosmological gas inflow.
In section \se{runs} we describe the sample of simulations. \se{Halpha} describes the mock \Halpha \ images of the sample.
\se{drops} discusses the metallicity drops found in the simulations and we connect them with recent gas accretion events in \se{infall}. Finally, \se{summary} is devoted to discussion and summary.

\section{Simulations}
\label{sec:runs}

 %
 The zoom-in simulations used in this paper were drawn from a larger dataset, the  \textsc{Vela} sample 
 \citep{Ceverino14, Zolotov}.
 We select for each simulation the latest snapshot that fulfills the two following criteria:
\begin{itemize}
\item
A galaxy stellar mass of $\Ms \le 10^9 \ \Msun$,
\item
A specific star-formation-rate of $sSFR \ge 1 \Gyr^{-1}$,
\end{itemize}
where $sSFR=SFR/\Ms$ (\tab{1}).
Therefore, we select star-forming galaxies with low masses ($\Ms=(0.4-1) \ 10^9 \ \Msun$), and gas rich (gas-over-star mass ratio about unity) at high-redshifts ($z=6.7-1.9$).
These protogalaxies are expected to undergo large, quasi-continuous gas inflows, so that they are the best candidates for setting the connection between cosmological gas inflow and metallicity drops.
Ongoing or recent major mergers with strong inflows of stars and dark matter 
could complicate this connection, so that they are excluded from the analysis.

\begin{table} 
\caption{Properties of the zoom-in simulations, ordered by decreasing gas fraction. Columns show name of the run, redshift, virial radius, virial mass, stellar mass, gas mass within 0.1$R_{\rm vir}$ and SFR in $\sy$. Masses are in $\Msun$ and radii in kpc.
Virial properties are computed using an overdensity, $\Delta$(z)$\simeq$180, above mean density \citep{Dekel13,BryanNorman98}.} 
 \begin{center} 
 \begin{tabular}{ccccccc} \hline 
\multicolumn{2}{c} {Run } \ \ \ z \ \ \ & $R_{\rm vir}$ & virial mass & stellar mass & gas mass & SFR \\
\hline 
V05   & 3.0 & 31 & $5.3 \times 10^{10}$ & $4.0  \times 10^8$ & $5.9  \times 10^8$ & 0.5 \\
V09   & 5.2 & 20 & $5.2 \times 10^{10}$ & $5.2  \times 10^8$ & $6.0  \times 10^8$ & 2.2 \\
V22   & 6.1 & 17 & $5.1 \times 10^{10}$ & $6.2  \times 10^8$ & $6.8  \times 10^8$ & 4.3 \\
V07   & 4.9 & 18 & $3.3 \times 10^{10}$ & $5.2  \times 10^8$ & $5.7  \times 10^8$ & 3.0 \\
V01   & 3.5 & 26 & $4.5 \times 10^{10}$ & $4.5  \times 10^8$ & $4.6  \times 10^8$ & 1.2 \\ 
V04   & 1.9 & 55 & $12 \times 10^{10}$ & $10  \times 10^8$ & $10  \times 10^8$ & 2.6 \\
V06   & 5.2 & 20 & $5.5 \times 10^{10}$ & $5.9  \times 10^8$ & $5.8  \times 10^8$ & 3.2 \\
V26   & 4.9 & 22 & $5.9 \times 10^{10}$ & $5.7  \times 10^8$ & $5.3  \times 10^8$ & 4.0 \\
V17   & 6.7 & 16 & $5.1 \times 10^{10}$ & $5.6  \times 10^8$ & $5.0  \times 10^8$ & 5.0 \\
V13   & 3.3 & 33 & $8.1 \times 10^{10}$ & $7.8  \times 10^8$ & $6.9  \times 10^8$ & 2.3 \\
V02   & 2.7 & 37 & $7.0 \times 10^{10}$ & $8.1  \times 10^8$ & $5.9  \times 10^8$ & 1.1 \\ 
V25   & 3.5 & 32 & $8.8 \times 10^{10}$ & $8.3  \times 10^8$ & $5.7  \times 10^8$ & 2.2 \\
V23   & 4.0 & 27 & $6.6 \times 10^{10}$ & $10  \times 10^8$ & $6.2  \times 10^8$ & 2.8 \\
V27   & 3.8 & 32 & $9.9 \times 10^{10}$ & $8.9  \times 10^8$ & $5.1  \times 10^8$ & 0.9 \\
V08   & 3.2 & 33 & $7.1 \times 10^{10}$ & $9.9  \times 10^8$ & $4.8  \times 10^8$ & 1.4 \\
V33   & 4.5 & 25 & $7.6 \times 10^{10}$ & $9.5  \times 10^8$ & $4.5  \times 10^8$ & 2.4 \\
V20   & 6.1 & 16 & $3.9 \times 10^{10}$ & $8.3  \times 10^8$ & $3.5  \times 10^8$ & 2.3 \\
 \end{tabular} 
 \end{center} 
\label{tab:1} 
 \end{table} 

%
The simulations were performed with the  \textsc{ART} code
\citep{Kravtsov97,Kravtsov03}, which accurately follows the evolution of a
gravitating N-body system and the Eulerian gas dynamics using an AMR approach.
Beyond gravity and hydrodynamics, the code incorporates 
many of the physical processes relevant for galaxy formation.  
These processes, representing subgrid 
physics, include gas cooling by atomic hydrogen and helium, metal and molecular 
hydrogen cooling, photoionization heating by a constant cosmological UV background with partial 
self-shielding, star formation and feedback, as described in 
\citet{Ceverino09}, \citet{CDB}, and \citet{Ceverino14}. 
 In addition to thermal-energy feedback, the simulations use radiative feedback.
This model adds a non-thermal pressure, radiation pressure, to the total gas pressure in regions where ionizing photons from massive stars are produced and trapped. 
In the current implementation, named RadPre in \citet{Ceverino14}, radiation pressure is included in the cells (and their closest neighbors) that contain stellar particles younger than 5 Myr and whose gas column density exceeds $10^{21}\ \cms.$ 
 
Metals are produced and advected according to the model described in \cite{Kravtsov03}.
The code follows two kinds of metals: metals produced in SN-II (Oxygen and other $\alpha$-elements), and metals produced in SN-Ia (Iron-like elements). 
Stars with masses between $m_* =8 \ \Msun$ and $100 \ \Msun$  eject $f_{\rm Z} \ m_*$ mass of metals in core-collapsed supernovae, where 
$f_{\rm Z}={\rm min}(0.2, 0.01 \ m_* - 0.06)$, which crudely approximates the SN yields of \cite{WoosleyWeaver95}.
After integrating a Chabrier IMF, this translates into the ejection of 
$\sim1\%$ of the initial stellar mass of a particle representing a single stellar population.
These metals are injected in the cell hosting the star particle at a constant rate during 40 Myr, 
the time that takes the last SN-II to explode. 
 
 The initial conditions of these runs contain from 6.4 to 46 $10^6$ dark matter particles 
 with a minimum mass of 
$8.3 \times 10^4 \ \msun$, while the particles representing single stellar populations that were formed in the simulation
have a minimum mass of $10^3 \ \msun$. 
The maximum spatial resolution is between 17-35 proper pc. More details can be found in \citet{Ceverino14} and \citet{Zolotov}.

\section{\Halpha\ Images}
\label{sec:Halpha}

 \begin{figure*} 
\includegraphics[width =0.99 \textwidth]{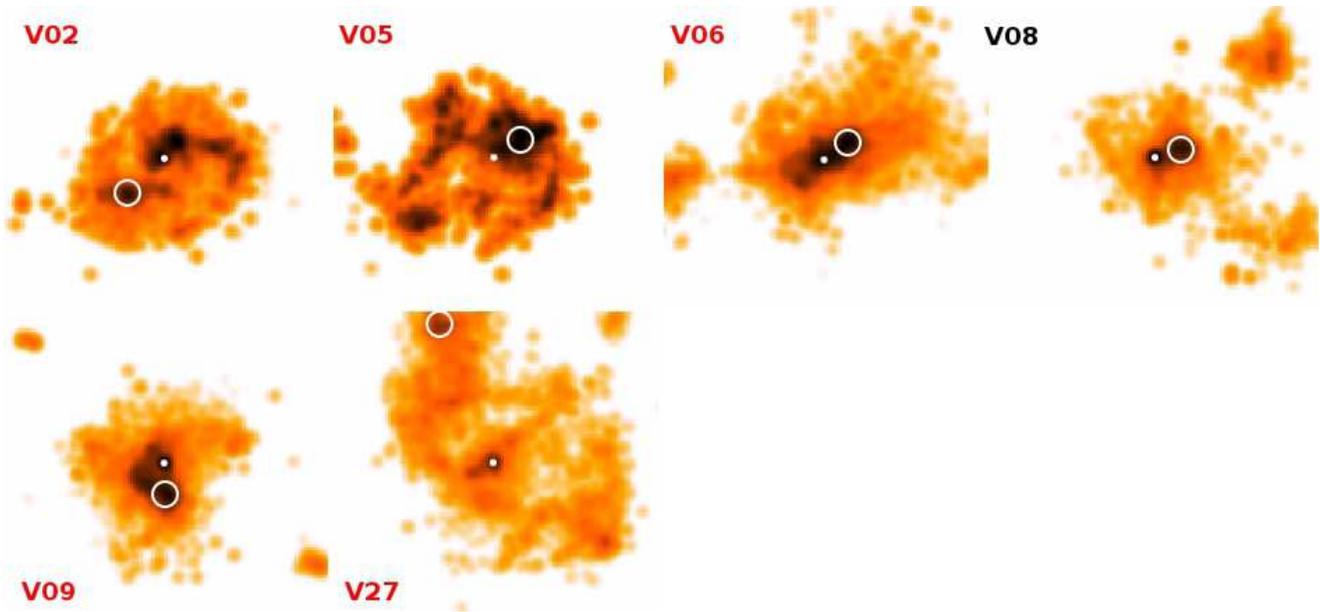}
\caption{Synthetic H$\alpha$ maps of galaxies with off-centre bright clumps, marked by a white circle with a radius of 0.4 kpc.
A white dot marks the galaxy center of mass.
The size of each image is $10\times 10$  kpc$^2$. In most of these runs, the clumps have a metallicity lower than in the inter-clump medium (red labels). A single case with a slightly off-center clump with metallicity higher than in the inter-clump is labeled in black. }
\label{fig:offcenter}
\end{figure*}

\begin{figure*} 
\includegraphics[width =0.99 \textwidth]{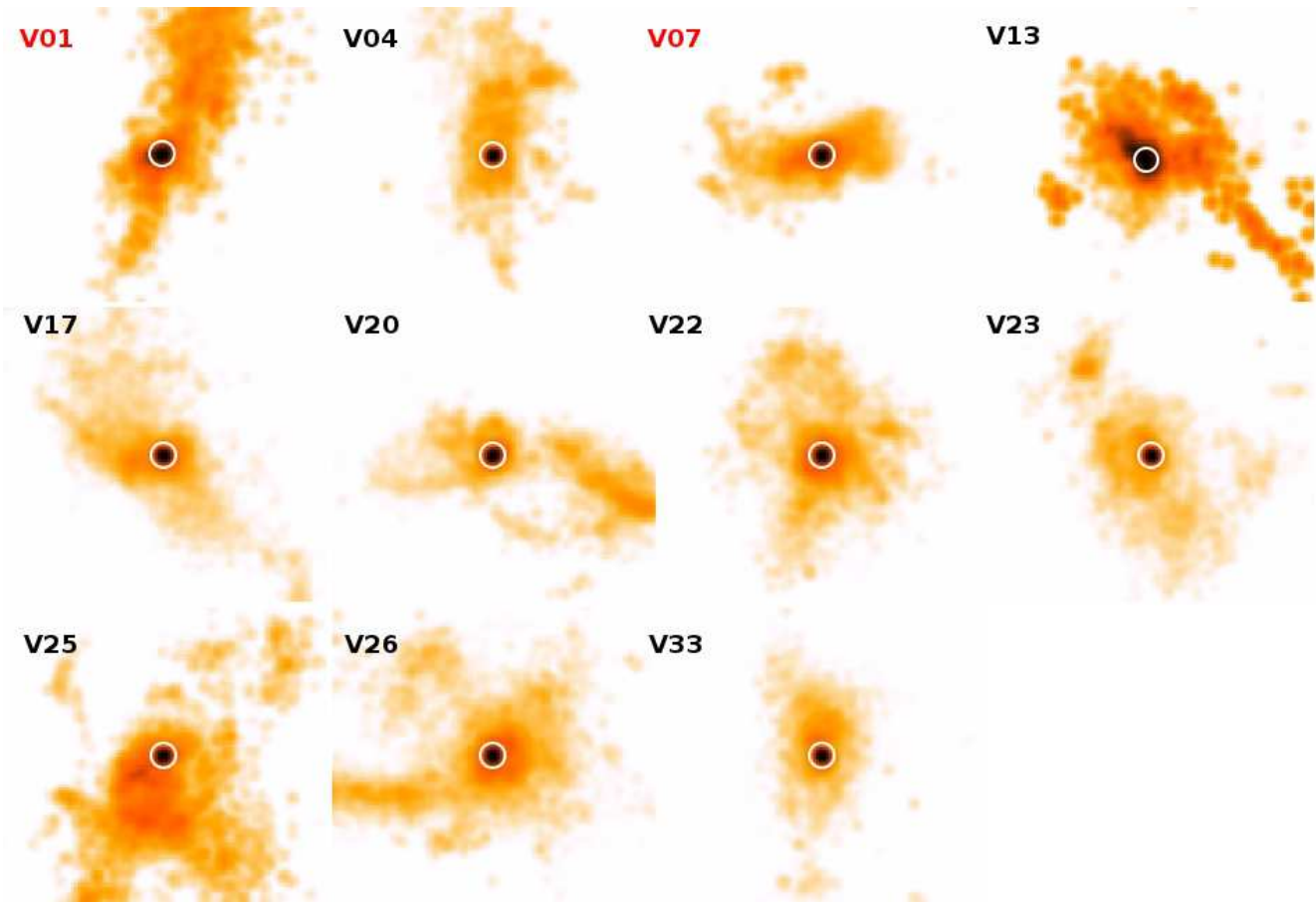}
\caption{Synthetic H$\alpha$ maps of galaxies with only one bright central clump marked by the white circle. 
The size of the images and the labels are as in \Fig{offcenter}.
2 out of 11 cases show metallicity drops at the galaxy centre.}
\label{fig:central}
\end{figure*}

If we want to make a connection between the gas inflow and the metallicity drops in starburst regions, the first step is to generate synthetic narrowband observations of \Halpha \ emission that highlights star-forming regions.
The mock \Halpha\ images of the sample   
are generated according to the method outlined in \citet{Ceverino12}.
In short, we first compute the star  
formation rate density, $\rho_{\rm SFR}$, using the distribution of stellar  
particles younger than 50 Myr.  
Then, we compute the \Halpha\ emissivity  
$\epsilon_{{\rm H}\alpha}$,  
based on the \citet{Kennicutt98} conversion, adjusted to a  
\citet{Chabrier03} IMF, 
\begin{equation} 
\log \epsilon_{\rm H\alpha} = \log \rho_{\rm SFR} + 41.33 \, , 
\end{equation} 
where $\epsilon_{\rm H\alpha}$ is in $\ergs \kpc^{-3}$  
and $\rho_{\rm SFR}$ is in $\sy \kpc^{-3}$. 
This equation holds as long as the \Halpha\ photons trace the underlying  
star-formation events, which is the case on scales of a few hundred parsecs  
\citep{Kennicutt07}. 
The \Halpha\ surface brightness is obtained by integrating the 
emissivity along a line-of-sight parallel to the galaxy rotation axis (face-on view).
Radiative transfer effects and dust extinction are not considered.
However, these effects are small in these dwarf galaxies, where the dust column density is low
\citep[][S{\'a}nchez Almeida et al., submitted]{Fisher14}. 
A Gaussian smoothing of FWHM=0.3 kpc is implemented in the final images.

The galaxies are classified in two groups, according to their morphology in the face-on, \Halpha\ images.
The first group looks clumpy in \Halpha, showing one or more \Halpha-bright, off-centre clumps (\Fig{offcenter}).
The second group shows only one  \Halpha-bright clump at the galaxy centre (\Fig{central}).
These \Halpha-bright clumps could be associated with the starburst regions observed in star-forming galaxies at high and low redshifts \citep{Genzel11, SA14a}.

Around one-third of the sample of star-forming galaxies are clumpy.
This clumpy fraction increases to $\sim$50\% if we exclude redshifts higher than 4. 
This is consistent with UV observations of bright clumps in high-z galaxies \citep{Guo15}. Their figure 10 shows a 60\% fraction of clumpy galaxies with stellar masses of log($\Ms / \Msun$)=9-10 in the redshift range $z=0.5-3$, slightly higher masses than in our sample,
(see also \citet{Elmegreen07} and \citet{Tadaki14}).

\section{Metallicity Drops}
\label{sec:drops}

\begin{figure} 
\includegraphics[width =0.49 \textwidth]{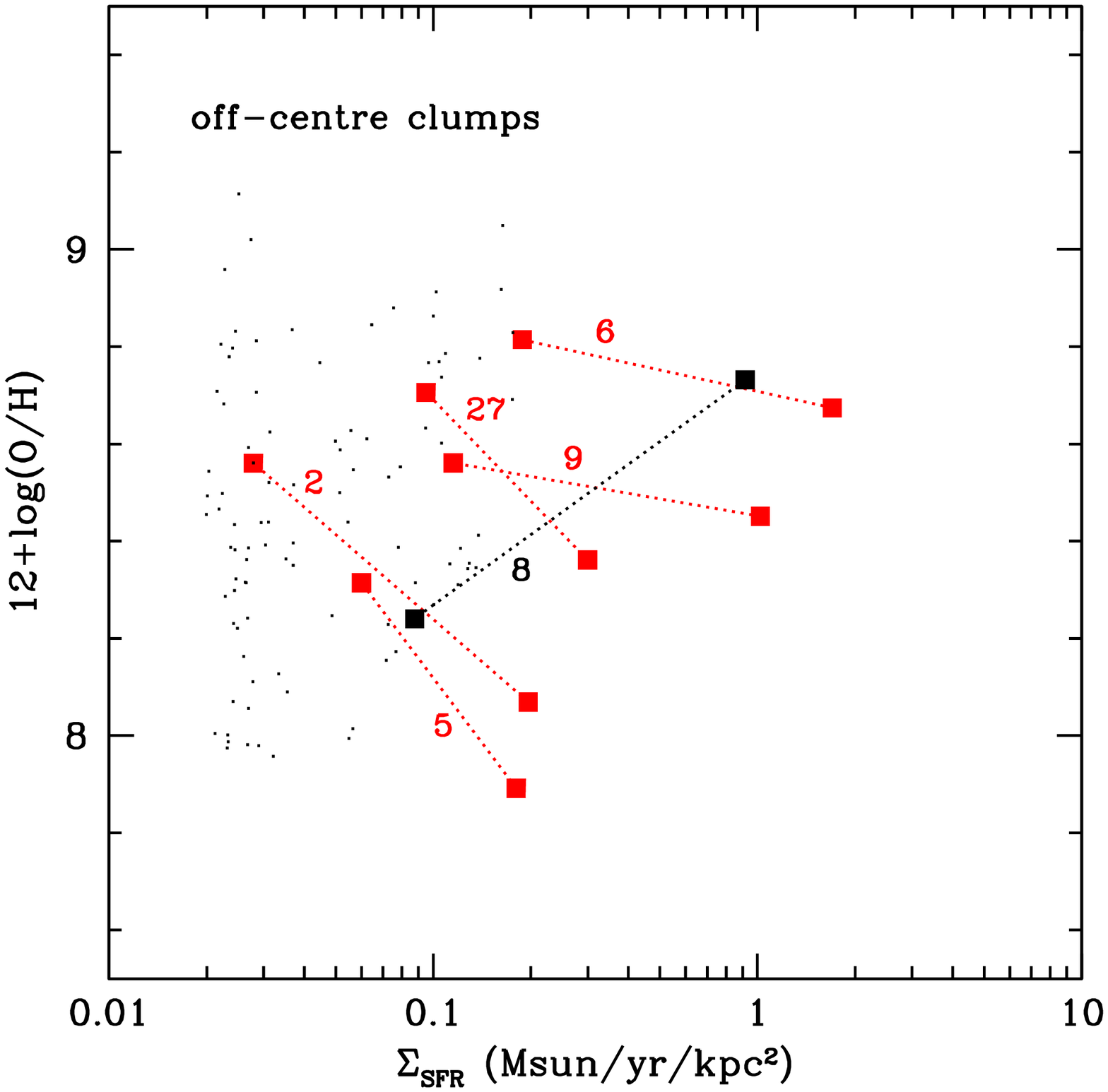}
\includegraphics[width =0.49 \textwidth]{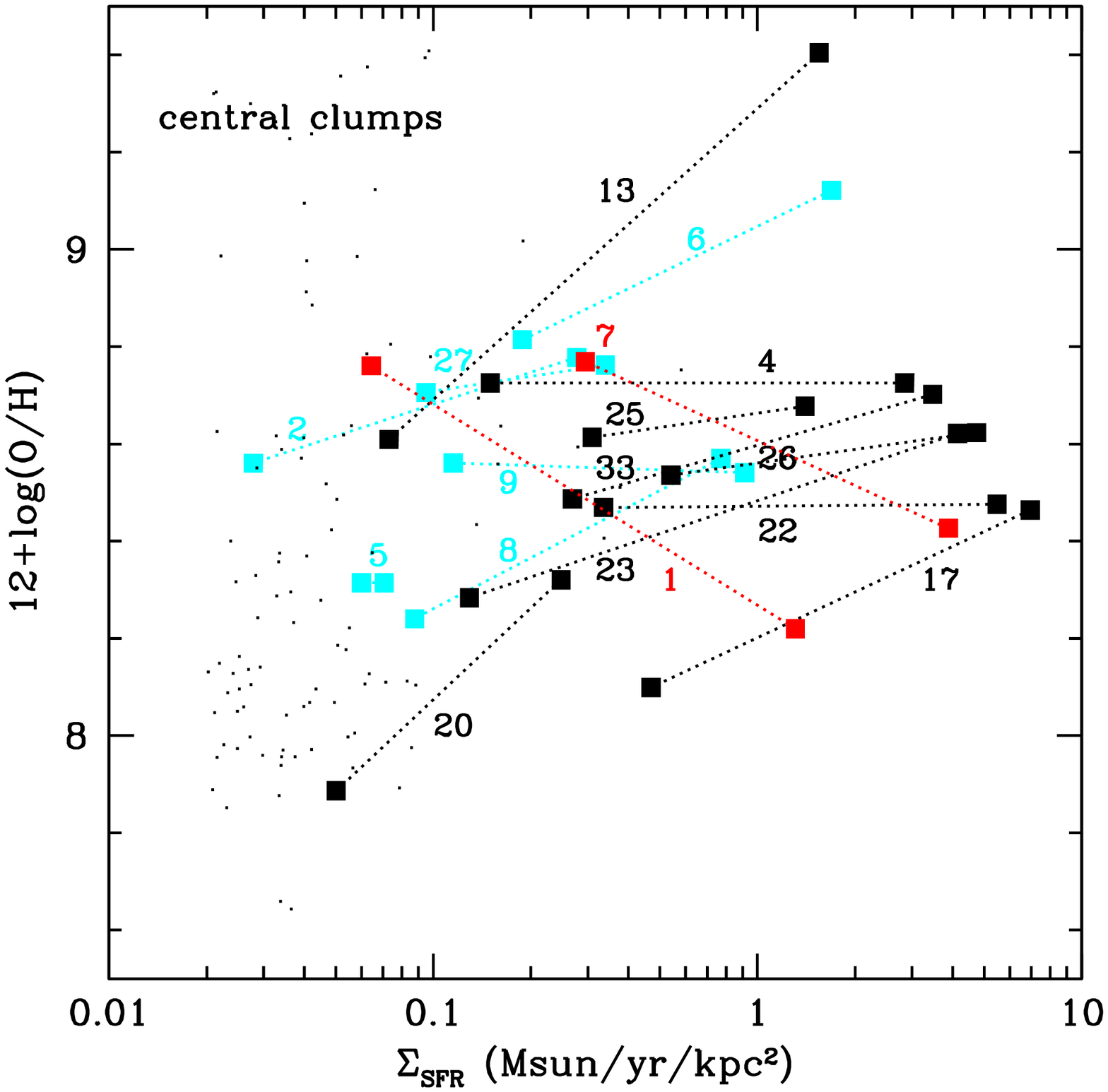}
\caption{Metallicity versus SFR surface density measured in apertures of 0.4 kpc radius. Each galaxy has two measures, linked by a dotted line, labeled by the run ID.
The highest SFR value in each galaxy corresponds to the \Halpha-bright clump. The lowest value corresponds to a randomly placed aperture.
The small dots represent 100 random apertures for V02 (top) and V13 (bottom), excluding the \Halpha-bright clump and the galaxy centre.
The squares are joined by red or black lines depending on whether the  clump is metal poorer or metal richer than the inter-clump medium.
Cyan squares and lines correspond to the center of clumpy galaxies.}
\label{fig:Zdiff}
\end{figure}

\begin{figure*} 
\includegraphics[width =0.99 \textwidth]{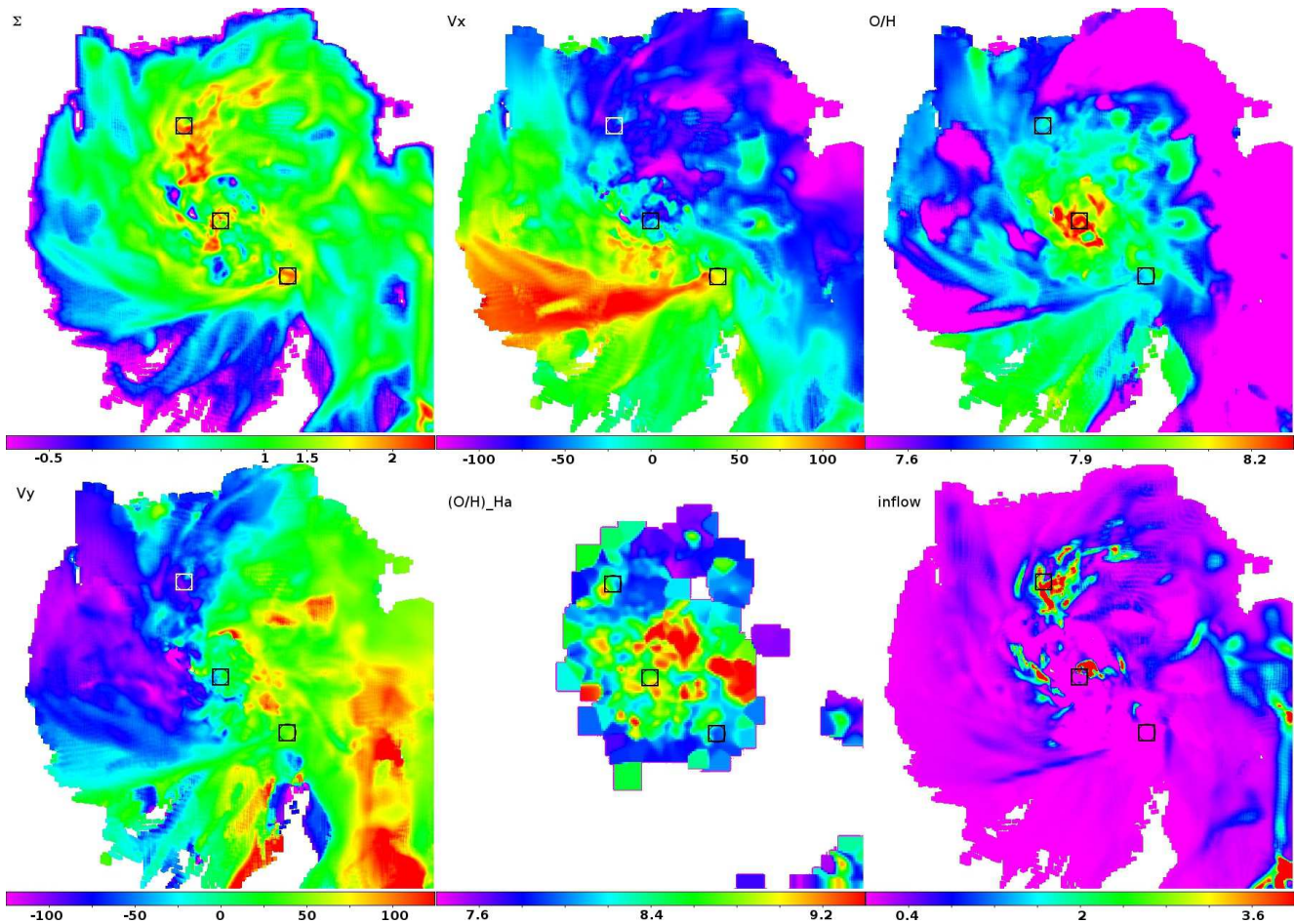}
\caption{Face-on maps of gas surface density ($\Sigma$), velocity in the X-axis of the galaxy plane ($V_X$), metallicity ($O/H$), velocity in the Y-axis ($V_Y$), gas metallicity weighted by H$\alpha$ emissivity ($O/H_{Ha}$), and radial inflow of V27 at $z=3.8$. In each panel, three square apertures mark the galaxy center and two regions with low metallicity  at the end of two incoming streams of pristine gas. Both apertures are close to H$\alpha$-bright star-forming regions. The size of the images is $ 20 \times 20$ kpc$^2$.}
\label{fig:vela27}
\end{figure*}

\begin{figure*} 
\includegraphics[width =0.99 \textwidth]{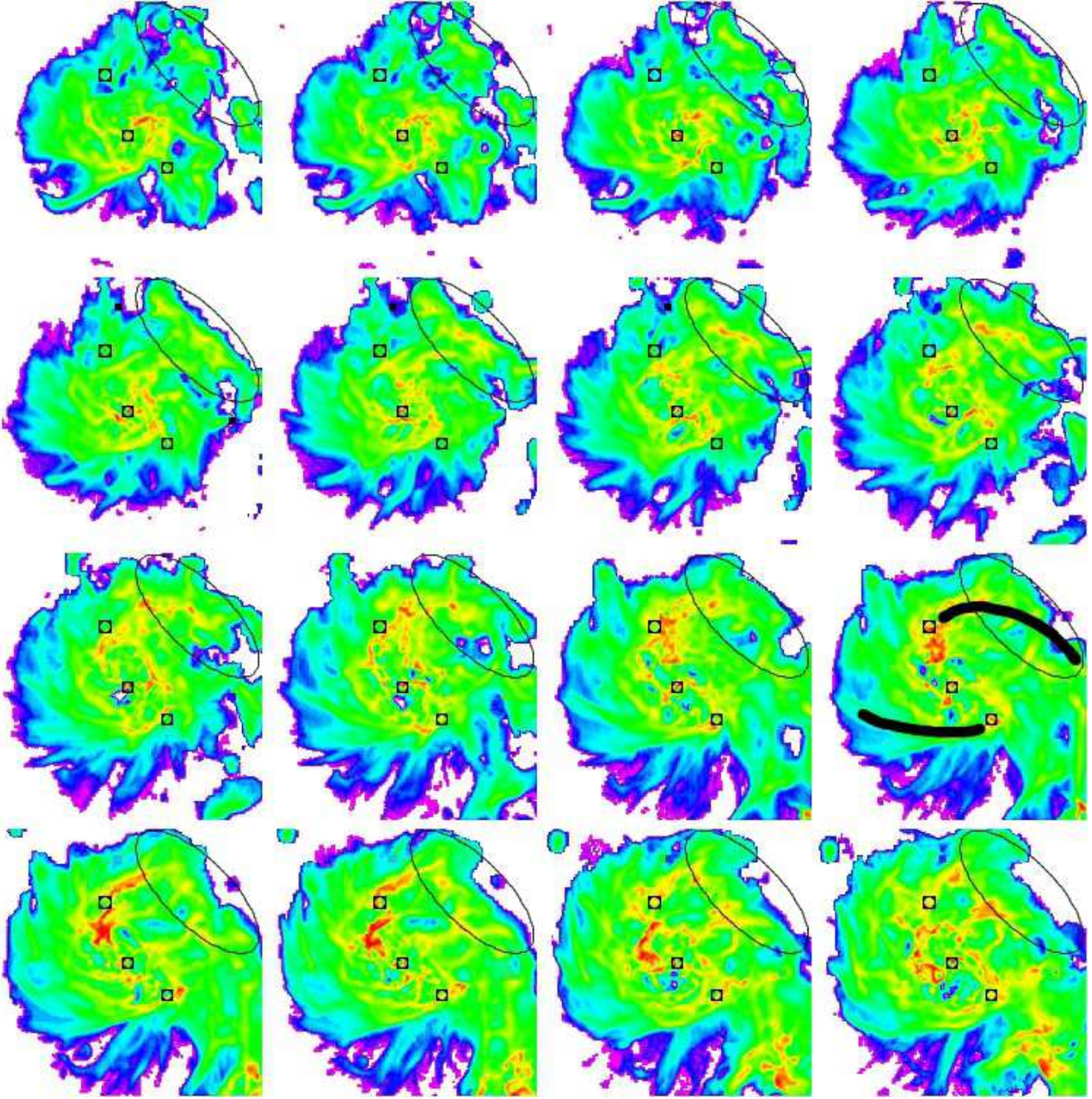}
\caption{Evolution of the gas surface density of V27, in time intervals of 10 Myr. Time increases from top-left to bottom-right. The rightmost panel of the third row (marked with the direction of the streams) coincides with the top-left panel of  \Fig{vela27}. 
The three black squares are located at the same positions as in \Fig{vela27}.
Dense, low-metallicity stream gas is coming from the right side of the view (marked by an ellipse) and accumulates material in the top-left side of the galaxy, increasing the gas surface density and the SFR in that region. 
These star-forming clumps are dispersed after $\sim$50 Myr,  i.e., a few times the disc dynamical time.
}
\label{fig:vela27_evo}
\end{figure*}

\begin{figure} 
\includegraphics[width =0.49 \textwidth]{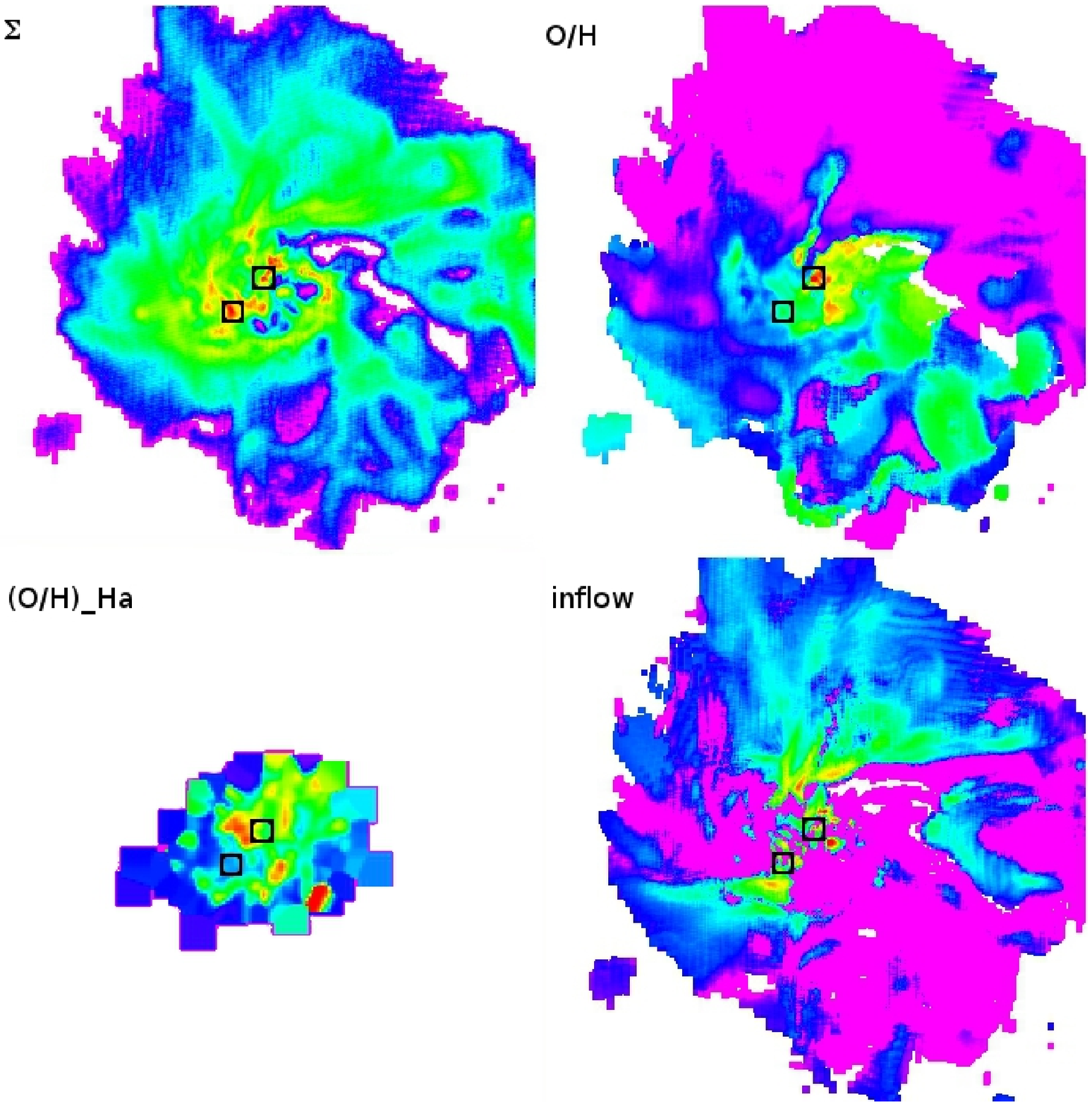}
\caption{Face-on maps of gas surface density ($\Sigma$), mass-weighted gas metallicity ($O/H$), gas metallicity weighted by H$\alpha$ emissivity ($O/H_{Ha}$), and radial inflow of V02 at $z=2.7$. In each panel, two square apertures mark the galaxy center and the off-centered clump. The size of the images is $ 20 \times 20$ kpc$^2$. The color palette is the same as in \Fig{vela27}.}
\label{fig:vela02}
\end{figure}

The next step is to find the metallicity drops by computing the mean gas metallicity, weighted by \Halpha \ emissivity, of the \Halpha-bright clumps and the inter-clump medium.
We translate the gas density of metals produced by supernova type-II  into oxygen abundances by using the formula described in \citet{Mandelker14}, 
\begin{equation} 
\frac{O}{H}  =   \frac{f_{\rm O}  \ z_{\rm SNII}}{ X  \ A } = \frac{0.5 \ z_{\rm SNII}}{ 0.755 \times  16 }
\end{equation} 
where $z_{\rm SNII}$ is the metal-to-gas mass ratio within a cell, 
$X=0.755$ is the hydrogen mass fraction, and $A=16$ is the atomic weight of oxygen.
We also assume that half of the mass in metals released in SNII are oxygen atoms ($f_{\rm O}=0.5$), roughly based on \citet{WoosleyWeaver95}.
The actual values of the scaling factors do not affect our results which are based on metallicity ratios rather than
on the absolute value of the metallicities.

The \Halpha-weighted gas metallicities are measured in circular apertures of radius 0.4 kpc, centered on 
the brightest off-center clump in \Halpha \ for each clumpy galaxy (\Fig{offcenter}).
For the non-clumpy galaxies (\Fig{central}), the aperture is centered at the \Halpha-bright clump at the galaxy center of mass.
In this first approach, we ignore all the smaller (dimmer) clumps, because they are usually not observed in the galaxies of this mass scale.
Therefore, we focus on the brightest clump that usually corresponds to the head in observed tadpole galaxies \citep{EE10,Elmegreen12}.
This is where a large amount of gas is needed to feed high SFRs.

In addition to the aperture centered on the clump, we also consider another aperture randomly placed in the inter-clump medium with a SFR surface density close to the galaxy median value. 
Therefore, each galaxy has two metallicity measures. 
The apertures centered on \Halpha-bright clumps have a high SFR surface density, much higher than in the inter-clump medium.
The actual position of  the aperture chosen for reference does not change the conclusions significantly,
as shown by 100 different random apertures for V02 and V13.

The top panel of \Fig{Zdiff} shows the metallicity in the clumpy galaxies of \Fig{offcenter}. 
The galaxies with metallicity drops are shown using red symbols and lines in \Fig{Zdiff}.
In almost all cases, the gas metallicity in the clump is significantly lower than the metallicity of the inter-clump medium. The median  drop in metallicity is 0.33 dex with a maximum value of 0.5 dex (V02).
The only case of an off-centered clump without a metallicity drop is shown in V08. 
This clump is also the nearest to the center of mass, so this could be an intermediate case between off-centered and central clumps.

The central clumps of \Fig{central} 
have higher SFR densities than the off-centered clumps, most probably because these galaxies are more compact than their clumpy counterparts.
The metallicity of the central clump is
 similar to or higher than the metallicity of the galaxy.
 More than 50\% of the random apertures described above have metallicities lower than the central value.
This is consistent with the fact that irregular galaxies with steep inner profiles could have negative metallicity gradients  \citep{Pilyugin15}.
V1 and V7 are two exceptional cases that show a central metallicity which is $\sim0.5$ dex lower than the metallicity at larger radii. In these cases, large radial flows of gas decrease the metallicity at the center.
 
It is remarkable that the  \Halpha-bright, off-centre clumps have lower metallicities 
than other star-forming regions with lower \Halpha \ luminosities,
because these clumps have high star formation rates and they are regions where significant amount of metals are injected into the gas due to supernova explosions.
These regions must have accreted large
amounts of gas with
low-metallicity, possibly coming from streams of inflowing gas. 
This inflow of low-metallicity gas could explain these inhomogeneities in metallicity.

\section{Gas inflow}
\label{sec:infall}

The inflowing gaseous streams join the disk through an outer ring, that can be tilted at large radii and then gradually aligned with the galaxy disk at low radii \citep{Danovich15}. This implies that the streams join the disk generally within a plane and at a given point in angle $\phi$, which may change as the streams-plus-galaxy system evolves.
\Fig{vela27} shows an example of gas accretion through streams (V27 at $z=3.8$).
Two streams coming from the circumgalactic medium can be seen at the left and right sides of the maps.
They are characterized by their low metallicities, high speed and inflow.
The streams meet the galaxy around the regions marked with two off-centered squares in \Fig{vela27}.
In these regions, gas accumulates and drives high SFR and high \Halpha \ luminosities (\Fig{offcenter}).
The \Halpha-weighted metallicity of these \Halpha-bright clumps is 
much lower than the average metallicity in the galaxy.
Therefore, these metallicity drops mark regions where low-metallicity streams accumulate material in the galaxy.

\Fig{vela27_evo} shows the temporal evolution of the gas in time steps of 10 Myr, 110 Myr before the snapshot shown in \Fig{vela27}. 
In this sequence, relatively dense low-metallicity gas is coming from the right side of the view. This stream of inflowing gas accumulates material in the top-left side of the galaxy, increasing the gas surface density and the SFR in that region.
Therefore,  gas accretion into galaxies promotes an \Halpha-bright burst of low-metallicity. 
After $\sim$50 Myr, a few times the disc dynamical time, these star-forming clumps dissolve due to the combination of radiative feedback \citep{Moody14} and turbulent motions  \citep{YangKrumholz12}.
The low metallicity drops are signatures or  hints of recent gas accretion into galaxies.

\Fig{vela02} shows another example of the link between metallicity drops and gas inflow.
In V02, the off-centered clump is much closer to the galaxy centre than in the previous case.
As shown in \Fig{vela27}, low-metallicity gas wraps around the galaxy.
The metal-poor clump, according to its \Halpha-weighted metallicity, is located at the end of one of the inflowing streams (at the bottom of the clump).
Other cases show similar patterns of gas accretion.

\begin{figure} 
\includegraphics[width =0.49 \textwidth]{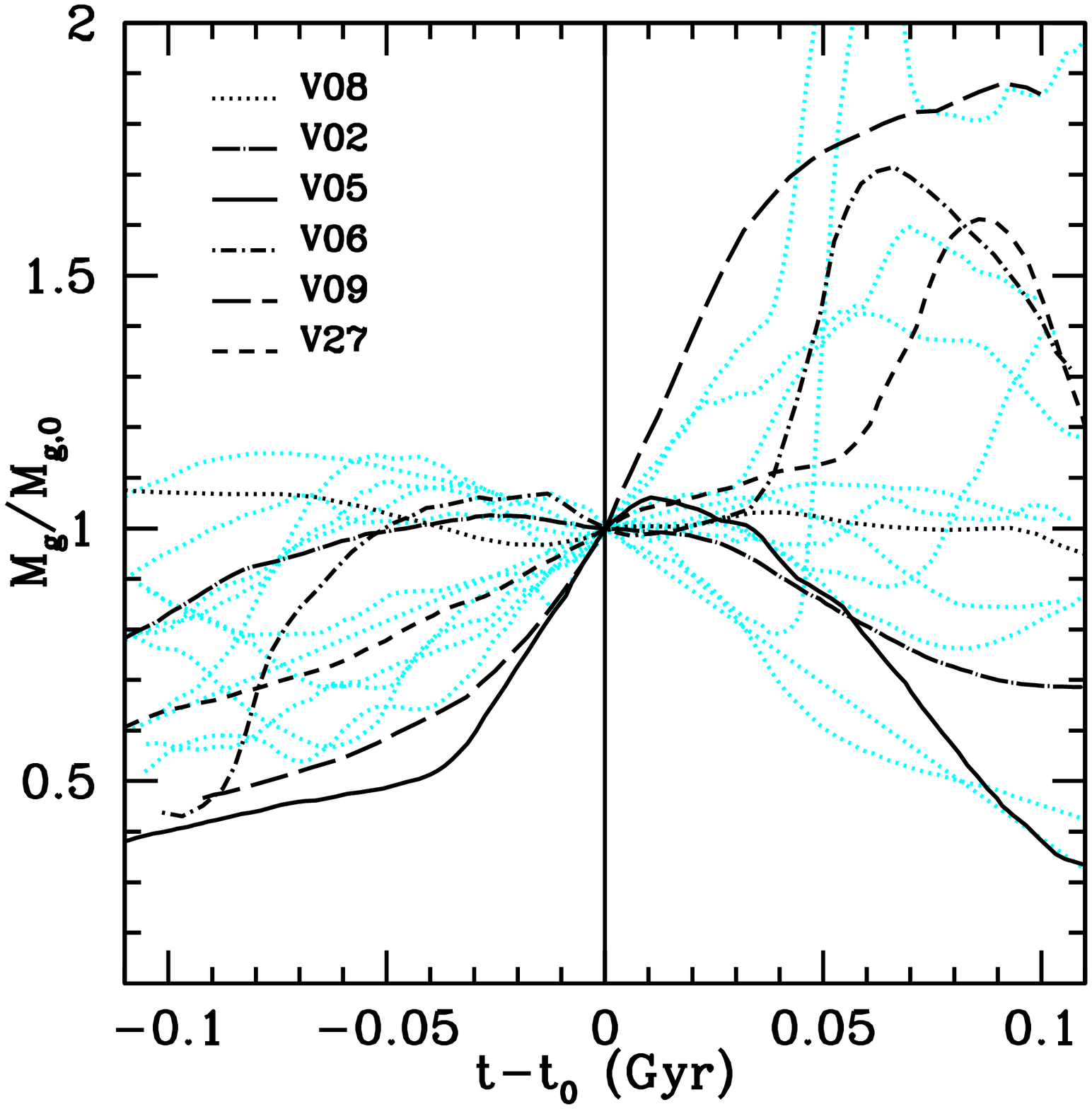}
\includegraphics[width =0.49 \textwidth]{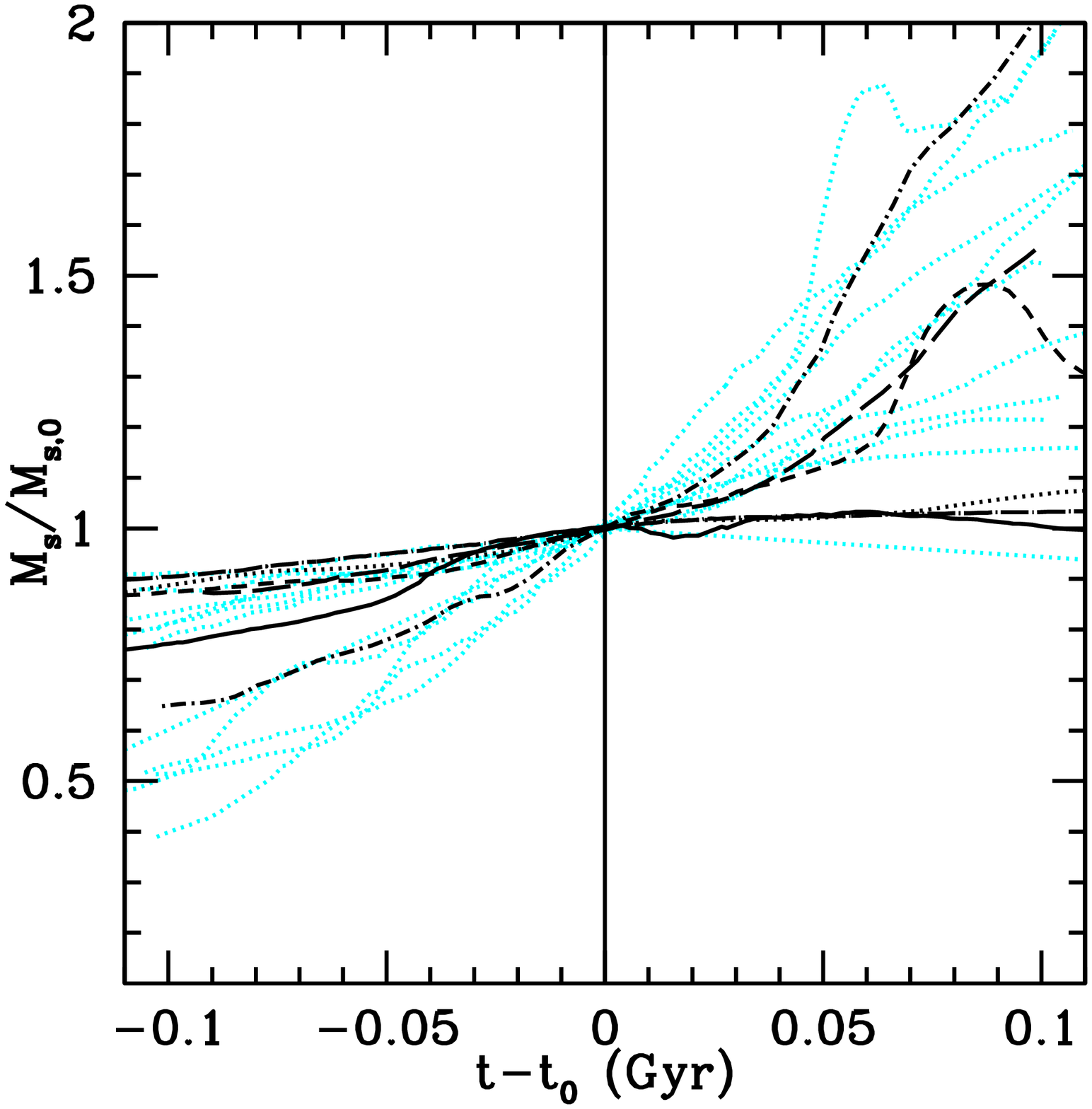}
\caption{Evolution of the gas mass (top) and stellar mass (bottom) within 0.15$R_{\rm vir}$ 
of the galaxies shown in \Fig{offcenter} and \Fig{central}.
 In all the five clumpy galaxies with metallicity drops induced by gas accretion (labeled), there is a significant increase in the amount of gas within $\sim$0.1 Gyr before the selected snapshots at $t_0$.
Cyan dotted lines correspond to the non-clumpy galaxies.
There is no significant stellar accretion with the exception of the mergers between 0.05 and 0.09 Gyr after $t_0$.}
\label{fig:infall}
\end{figure}

We can quantify this gas accretion by following the evolution of the gas mass within a sphere of fixed radius, 0.15$R_{\rm vir}$. \Fig{infall} shows this temporal evolution for gas and stars of the 
clumpy and non-clumpy galaxies shown in \Fig{offcenter} and \Fig{central}. 
Masses are normalized to the value at the selected snapshot (\tab{1}). 
We choose a time interval of $\pm 0.1 \Gyr$ around the snapshot, so we can quantify the inflow (or outflow) before (and after) that moment captured in \Fig{offcenter}
and we allow some time for the formation of clumps from this inflow material.
 
In three out of five cases with metallicity drops, there is a 
steep increase in the gas mass  in a period of 0.1 Gyr before the selected snapshot.
The gas mass doubles within this period. 
The increase in gas  is not related with merger events, 
because there is no sudden increase of the stellar mass within the same period. 
Instead, the stellar mass increases slowly, about 10-20\%, due to star-formation within the selected sphere.
A exception is a merger event in V27 at 0.08 Gyr.
Therefore, the increase of gas mass in these galaxies with metallicity drops is due to the continuous infall of gas from cosmic streams.

\section{Summary and Discussion}
\label{sec:summary}

We addressed the continuous gas accretion of low metallicity gas into star-forming galaxies at different redshifts, 
 using a set of zoom-in AMR cosmological simulations of galaxy formation.
In $\sim$50\% of the cases at redshifts lower than 4, the accretion gives rise to star-forming, \Halpha-bright, off-centre clumps.
Most of these clumps have gas metallicities, weighted by \Halpha \ luminosity, lower than the average metallicity in the surrounding interstellar medium. The typical values of these metallicity drops are around 0.3-0.5 dex, consistent with observations \citep{Cresci10, SA13, SA14a,SA15}.
Therefore, the observed metallicity drops can be considered as evidence for rapid gas accretion coming from cosmological inflow of low-metallicity gas. 

Low metallicity inhomogeneities are dispersed in $\sim$50 Myr, only a few times the disc dynamical time, 
$t_{\rm d}=r_{\rm d} / v_{\rm d}= 2 \kpc / 100 \kms = 20$ Myr, where $r_{\rm d}$ and $v_{\rm d}$ are the characteristic radius and rotational velocity of these galaxies.
Therefore, low-metallicity gas gets mixed with galactic interstellar medium in less than an orbital period.
These short time-scales are consistent with the time-scales for dispersion of non-axisymmetric inhomogeneities in galaxy discs \citep{PetitKrumholz15}.
The metal mixing is led by shear and turbulence, driven by gravitational disc instabilities.
Shear accelerates the diffusion of metals by transferring inhomogeneities from large scales to small scales, where turbulence is able to dissipate them in a dynamical time-scale \citep{YangKrumholz12}. 
Feedback may also contribute to the dispersion of metals by the disruption of the small metal-poor clumps \citep{Moody14}.
More massive galaxies hosting giant, long-lived clumps \citep{DSC,CDB,Mandelker14,Bournaud14}, may have metallicity drops with longer lifetimes, if clump self-enrichment due to supernovae yields is not a relevant process. Future work focused on these giant clumps will clarify these issues.


\section*{Acknowledgments} 
 
The simulations were performed 
at the National Energy Research Scientific Computing Center (NERSC) at  
Lawrence Berkeley National Laboratory, and 
at NASA Advanced Supercomputing (NAS) at NASA Ames Research Center.
This work has been partly funded by the Spanish Ministry of 
 Economy and Competitiveness, projects  AYA2012-32295 and
 AYA2013--47742--C04--02--P.

\bibliographystyle{mn2e}
\bibliography{drops6}

\bsp

\label{lastpage}

\end{document}